\documentclass[12pt,onecolumn,showpacs,superscriptaddress,apjl,amsmath,amssymb,freqn]{emulateapj}
\usepackage{color}
\usepackage{graphicx}
\usepackage{amssymb,amsmath,epsfig}
\usepackage{natbib}
\input{colordvi.tex}

\numberwithin{equation}{section} 

\newcommand{\kvec}{{\boldsymbol{k}}}
\newcommand{\uvec}{{\boldsymbol{u}}}
\newcommand{\nhat}{{\hat{\boldsymbol{n}}}}

\newcommand{\uvecsix}{{\uvec_1 ,\uvec_2 ,\uvec_3 ,\uvec_4 ,\uvec_5 ,\uvec_6}}

\begin{document}

\title{Sensitivity to 21cm Bispectrum from Epoch of Reionization}

\author{Shintaro Yoshiura$^1$, Hayato Shimabukuro$^{1,2}$, Keitaro Takahashi$^1$, Rieko Momose$^3$, Hiroyuki Nakanishi$^4$, and Hiroshi Imai$^4$}
\email{}

\altaffiltext{1}{Department of Physics, Kumamoto University, Kumamoto,Japan}
\altaffiltext{2}{Department of Physics, Nagoya University, Aichi , Japan}
\altaffiltext{3}{Institute for Cosmic Ray Reaserch, Tokyo University,Tokyo,Japan}
\altaffiltext{4}{Department of Physics, Graduate School of Science and Engineering, Kagoshima University,Kagoshima,Japan}

\begin{abstract}
The 21cm line brightness temperature brings rich information about Epoch of Reionizaton (EoR) and high-$z$ universe (Cosmic Dawn and Dark Age). While the power spectrum is a useful tool to investigate the EoR signal statistically, higher-order statistics such as bispectrum are also valuable because the EoR signal is expected to be highly non-Gaussian. In this paper, we develop a formalism to calculate the bispectrum contributed from the thermal noise taking array configularion of telescopes into account, by extending a formalism for the power spectrum. We apply our formalism to the ongoing and future telescopes such as expanded Murchison Widefield Array (MWA), LOw Frequency ARray (LOFAR) , Hydrogen Epoch of Reionization Array (HERA) and Square Kilometre Array (SKA). We find that expanded MWA does not have enough sensitivity to detect the bispectrum signal. On the other hand, LOFAR has better sensitivity and will be able to detect the peaks of the bispectrum as a function of redshift at large scales with comoving wavenumber $k \lesssim 0.03~{\rm Mpc}^{-1}$. The SKA has enough sensitivity to detect the bispectrum for much smaller scales $k \lesssim 0.3~{\rm Mpc}^{-1}$ and redshift $z \lesssim 20$
\end{abstract}

\keywords{cosmology: observations, dark ages, reionization, first stars. instrumentation: interferometers. techniques: interferometric}

\section{Introduction}

The redshifted 21cm line emission from neutral hydrogens is a promising way to probe Epoch of Reionization (EoR), Cosmic Dawn and Dark Age \citep{2006PhR...433..181F,2012RPPh...75h6901P} because it reflects the physical state of intergalactic gas. Actually, the brightness temperature depends on quantities crucial for the understanding of these epochs, such the neutral hydrogen fraction, spin temperature and baryon density. However, the observation of the redshifted 21cm signal is very challenging due to the presence Galactic and extragalactic foreground emissions, Low-frequency radio telescopes such as Murchison Widefield Array (MWA) \citep{2013MNRAS.429L...5B}, LOw Frequency ARray (LOFAR) \citep{2013MNRAS.435..460J} and PAPER \citep{2014ApJ...788..106P} have started their observations and set upper bounds on the brightness temperature. The upper bounds will improve further as our understanding of the foreground proceeds and the subtraction techniques become more sophisticated. Ultimately, the Square Kilometre Array (SKA) \citep{SKA,2013ExA....36..235M} and Hydrogen Epoch of Reionization Array (HERA) \citep{2014ApJ...782...66P} will perform precise observations and will reveal the physical process of EoR and Cosmic Dawn.

One of the useful tools to extract information from observed data is to take the power spectrum of fluctuations in brightness temperature at a fixed redshift (frequency). This is effective even for relatively low S/N data, which could be obtained by ongoing telescopes, while making a map of brightness temperature through imaging requires much higher sensitivity the SKA is expected to have. Actually, the power spectrum of brightness temperature has been studied by many authors \citep{2007MNRAS.376.1680P,2014ApJ...782...66P,shi-P,2006PhR...433..181F,2010A&A...523A...4B,2014MNRAS.439.3262M,2008ApJ...689....1S}.

When fluctuations follow Gaussian probability distribution, they can be well characterized by the power spectrum and higher-order statistics such as bispectrum and trispectrum have no further independent information. However, since reionization is a highly non-Gaussian process which involves non-linear density fluctuations, star formation and expansion of HII bubbles, the brightness temperature fluctuations are also expected to be strongly non-Gaussian \citep{shi-P}. In this case, the power spectrum does not have sufficient information to describe the fluctuations and higher-order statistics have independent and complimentary information \citep{2005MNRAS.363.1049C,2007ApJ...662....1P}.

In this paper, we develop a formalism to calculate the errors in bispectrum measurement contributed from thermal noise. Noise estimation has been studied by many authors in case of power spectrum \citep{2004ApJ...615....7M,2005ApJ...619..678M,2006ApJ...653..815M}, and we extend the formalism given in \cite{2006ApJ...653..815M}. Starting from the error in visibility obtained by a single baseline, we consider its summation over the baseline distribution in $uv$ plane. A striking feature of thermal-noise bispectrum is that its ensemble average vanishes because thermal noise is Gaussian. Nevertheless, thermal noise contributes to the bispectrum error through its variance. Considering the variance of thermal noise error is the main extention to the previous formalism.

The structure of this paper is the following. In section 2, we define the brightness temperature, it's power spectrum and bispectrum. In section 3, we review the formalism of calculation of thermal-noise power spectrum given by \cite{2006ApJ...653..815M}. Then, we develop a formalism for bispectrum and estimate thermal-noise bispectrum for several specific configuration of the wave number in section 4. The summary and discussion will be given in section 5. Throughout this paper, we assume $\Lambda$CDM cosmology with $(\Omega_{\rm m},\Omega_\Lambda,\Omega_{\rm b},H_0) = (0.27,0.73,0.046,70~{\rm km/s/Mpc})$ \citep{2011ApJS..192...18K}.

\section{21cm line signal}

In this section, we define basic quantities concerning the 21cm signal. The brightness temperature $\delta T_b$ is defined by spin temperature offsetting from CMB temperature,
\begin{eqnarray}
\delta T_b(z)
&=& \frac{T_s - T_\gamma}{1+z}(1 - e^{-\tau_{\nu_{0}}}) \nonumber\\
&\approx& 27 x_{\rm HI} (1 + \delta_m)
          \bigg(\frac{H}{dv_r/dr + H} \bigg)
          \bigg(1 - \frac{T_\gamma}{T_{\rm S}}\bigg)
          \bigg(\frac{1+z}{10} \frac{0.15}{\Omega_m h^2}\bigg)^{1/2}
          \bigg(\frac{\Omega_b h^2}{0.023} \bigg) ~ [{\rm mK}],
\label{eq:brightness}
\end{eqnarray}
where $x_{HI}$ is the neutral fraction of hydrogen, $\delta_m$ is the matter over density, $H$ is the Hubble parameter and $dv_r/dr$ is the velocity gradient along the line of sight. Then we introduce fluctuation of $\delta T_b({\bf x})$,
\begin{equation}
\delta_{21}({\bf x}) = \delta T_b({\bf x}) - \delta \bar{T_b},
\label{delta_T_b}
\end{equation}
where $\delta \bar{T_b}$ is the average value of brightness temperature, ${\bf{x}}$ is spatial position. The power spectrum of brightness temperature is defined from its Fourier transform, $\delta_{21}(\kvec)$, as,
\begin{equation}
\langle \delta_{21}({\bold k_1}) \delta_{21}({\bold k_2}) \rangle
= \delta({\bold k_1} + {\bold k_2}) P_{21}(k_1),
\label{eq:ps_def}
\end{equation}
where $\langle \rangle$ represents the ensemble average, ${\bf{k}}$ is position in Fourier space. The bispectrum $B_{21}$ can be defined in a similar way:
\begin{equation}
\langle \delta_{21}({\bold k_1}) \delta_{21}({\bold k_2}) \delta_{21}({\bold k_3}) \rangle
= \delta({\bold k_1}+{\bold k_2}+{\bold k_3})
  B_{21}({\bold k_1},{\bold k_2}). 
\label{eq:bs_def}
\end{equation}
Here the delta function forces the three wave vectors to make a triangle and $B_{21}$ is dependent on only two of the three vectors (chosen $\bold k_1$ and $\bold k_2$ here) due to this triangle condition.

\section{power spectrum sensitivity}

In this section, we summarize a formalism to estimate the thermal noise for power spectrum, following \cite{2006ApJ...653..815M}. First, we define visibility $V(u,v,\nu)$ for a pair of antennae as,
\begin{equation}
V(u,v,\nu) =
\int{d{\nhat} \, \, T_N(\nhat,\nu) \, W(\nhat,\nu)
     e^{2 \pi {\it i}{u \choose v} \cdot \nhat}},
\label{viseqn}
\end{equation}
where $T_N$ is the thermal-noise temperature, $\nhat$ is the direction of primely beam, $\nu$ is observed frequency and $W(\nhat,\nu)$ is a product of the window functions concering the field of view and bandwidth. The rms thermal-noise fluctuation per visibility is given by,
\begin{eqnarray}
V_N = \frac{\lambda^2 T_{\rm sys}}{A_e \sqrt{\Delta \nu t_0}} ~ [\rm K],
\label{visnoise}
\end{eqnarray}
where $\lambda$ is the observed wavelength, $T_{\rm sys}$ is the total system temperature, $A_e$ is the effective area of antenna, $\Delta \nu$ is the width of the frequency channel and $t_0$ is total observing time. By Fourier transforming the visibility in the frequency direction, we obtain,
\begin{eqnarray}
\tilde{I}(u,v,\eta)
&=& \int d\nu V_N(u,v,\nu) \exp(2\pi i \nu \eta) \nonumber\\
&=& \sum^{B/\Delta \nu}_{i=1} V_N(u,v,\nu_i) \exp(2 \pi i \nu_i \eta)
    \Delta\nu ~ [\rm K \cdot Hz],
\end{eqnarray}
where $B (\gg \Delta \nu)$ is the bandwidth, $\nu_i$ is the $i$-th frequency channel and we define $\uvec = (u,v,\eta)$.

The covariance matrix of detector noise for a single baseline is given by,
\begin{eqnarray}
C_N(\uvec_i,\uvec_j)
&=& \langle \tilde{I}_N(\uvec_i) \tilde{I}^{*}_N(\uvec_j) \rangle \nonumber \\
&=& \int d\uvec' \int d\uvec''
    \langle \tilde{T}_N(\uvec') \tilde{T}_N(\uvec'') \rangle
    \tilde{W}(\uvec_i - \uvec') \tilde{W}(\uvec_j - \uvec'') \nonumber\\
&=& \int d\uvec' \int d\uvec''
    P_N(\uvec') \delta^3_D(\uvec' - \uvec'')
    \tilde{W}(\uvec_i - \uvec') \tilde{W}(\uvec_j - \uvec'') \nonumber\\
&=& \int d\uvec' \int d\uvec'' P_N(\uvec')
    \tilde{W}(\uvec_i - \uvec') \tilde{W}(\uvec_j - \uvec') \nonumber\\
&\approx& \delta_{ij} P_N(\uvec_i)
          \int d^3\uvec' \, |\tilde{W}(\uvec_i - \uvec')|^2,
\end{eqnarray}
where, we used the definition of power spectrum for noise temperature (Eq. \ref{eq:ps_def}) in third equality and assumed that the covariance vanishes when $\uvec_i \neq \uvec_j$ in the last equality. Further, we assumed that the power spectrum is constant for a range where the window function have non-zero value and we pulled $P_N$ out of the integration. Then the integration of window functions can be evaluated as follows:
\begin{eqnarray}
\int d^3\uvec' \, |\tilde{W}(\uvec - \uvec')|^2
&=& \int d^3\uvec' \int d^3{\bf r} \int d^3{\bf r'} \,
    |{W}({\bf r})| |{W}({\bf r'})|
    e^{2\pi i(\uvec - \uvec') \cdot (\bf{r} + \bf{r'})}\nonumber\\
&=& \int d^3{\bf{r_0}} \int d^3 {\bf r' } \delta_{D}({\bf r_0}) \,
    |{W}({\bf{r_0 - r'}})| |{W}({\bf{r'}})|
    e^{2 \pi i \uvec \cdot ({\bf r}_0)} \nonumber\\
&=& \int d^3{\bf r'} \, |{W}({\bf r'})| |{W}({\bf -r'})|
\approx \Omega B
\approx \frac{\lambda^2 B}{A_e},
\label{window}
\end{eqnarray}
where $\Omega$ is the field of view. Thus we obtain,
\begin{equation}
C_N(\uvec_i,\uvec_j) \approx \frac{\lambda^2 B}{A_e} P_N(\uvec_i) \delta_{ij}.
\label{eq:cn}
\end{equation}

On the other hand, the covariance matrix for a single baseline can be evaluated from Eq. (\ref{visnoise}),
\begin{eqnarray}
C_{N,1b}(\uvec_i,\uvec_j)
&=& \langle \tilde{I}(\uvec_i) \tilde{I}^*(\uvec_j) \rangle_{1b}
 =  \sum_l^{B/\Delta \nu} \sum_m^{B/\Delta\nu}
    |V_N(u_i,v_i,\nu_l) \Delta \nu |^2
    \delta_{ij} \delta_{lm} \nonumber\\
&=& \sum_l^{B/\Delta\nu} |V_N(u_i,v_i,\nu_l)|^2 \Delta \nu ~ \delta_{ij}
 =  \frac{B}{\Delta \nu} (\Delta \nu)^2 ({V_N}(u_i,v_i,\nu))^2 \delta_{ij} \nonumber\\
&=& \Bigl(\frac{\lambda^2 B T_{\rm sys}}{A_e}\Bigr)^2
    \frac{\delta_{ij}}{B t_0}.
\label{eq2}
\end{eqnarray}
Again, we assumed that there is no correlation between the thermal noise with different $u, v$ and $\nu$. If multiple baselines contribute to the same pixel, the observing time is effectively increased. Here we assume that the number density of the baselines in $uv$-plane is constant under rotation with respect to $\eta$-axis, that is, depends only on $|\uvec_\perp| = |\uvec| \sin{\theta}$ where $\theta$ is the angle between $\uvec$ and $\eta$-axis. Therefore, the effective observing time $t_{\uvec}$ can be written as,
\begin{equation}
t_\uvec \approx \frac{A_e}{\lambda^2} n(|\uvec| \sin{\theta}) t_0.
\label{eq:tu}
\end{equation}
Here $A_e/\lambda^2$ represents area per pixel on $uv$-plane which reflects the resolution on $uv$-plane and $n(|\uvec| \sin{\theta})$ is the number density of baselines on $uv$-plane. Thus, we obtain the covariance matrix for a pixel in $uv\eta$-space, replacing $t_0$ with $t_k$, as,
\begin{eqnarray}
C_N(\uvec_i,\uvec_j) = 
\Bigl(\frac{\lambda^2 B T_{\rm sys}}{A_e}\Bigr)^2 \frac{\delta_{ij}}{B t_\uvec}.
\label{eq:noise_pix}
\end{eqnarray}
Thus, comparing with Eq. (\ref{eq:cn}) and substituting Eq. (\ref{eq:tu}), we obtain,
\begin{equation}
P_N(\uvec) = \frac{\lambda^4 T_{\rm sys}^2}{A_e^2 n(|\uvec| \sin{\theta}) t_0}.
\end{equation}

Now we convert the noise power spectrum of $\uvec$ space to the one of cosmological Fourier space $\kvec$. Using the following relations
\begin{eqnarray}
&& \uvec_{\perp}
   = \frac{D_M(z)}{2 \pi} \kvec_{\perp} 
   \equiv \frac{x}{2 \pi} \kvec_{\perp},  \\
&& \eta
   \approx \frac{c (1+z)^2}{2 \pi H_0 f_{21}E(z)} k_z
   \equiv \frac{y}{2 \pi B} k_z,
\end{eqnarray}
where, $H_0$ is the Hubble constant, $f_{21}$ is the frequency of 21cm radiation and
\begin{eqnarray}
&& D_M(z) = \frac{c}{H_0} \int^z_0 \frac{dz'}{E(z')} \\
&& E(z) = \sqrt{\Omega_{\rm M} (1+z)^3 + 1 - \Omega_{\rm M}}
\end{eqnarray}
where $\Omega_{\rm M}$ is the density parameter of matter and we assumed the flat universe. Thus, we obtain,
\begin{equation}
P_N(\kvec) = \frac{x^2 y}{B} P_N(\uvec)
= \frac{x^2 y \lambda^4 T_{\rm sys}^2}{B A_e^2 n(|\uvec| \sin{\theta}) t_0}.
\label{eq:Ppix}
\end{equation}

Because the power spectrum of 21cm signal is dependent only on the length of the wave vector, we take a sum of the above noise power spectrum over a spherical shell which corresponds to the same $k$. First, we consider an annulus with radial width $\Delta k$ and angular width $\Delta \theta$. Noting that the baseline distribution is assumed to be uniform in an annulus, the number of pixels in the annulus is,
\begin{eqnarray}
N_a = 2 \pi k^2 \sin{\theta} ~ \Delta \theta ~ \Delta k \frac{V}{(2\pi)^3},
\label{pixnum}
\end{eqnarray}
where $V = \lambda^2 x^2 y/A_e$ is the observed volume in real space, $(2\pi)^3 / V$ is the resolution in Fourier space and the other factor, $2\pi k^2 \sin{\theta} \Delta \theta \Delta k$, is the annulus volume in Fourier space. Then the noise power spectrum ruduces by a factor of $1/\sqrt{N_a}$. Next, we consider a sum over $\theta$. Taking $\Delta k = \epsilon k$, where $\epsilon$ is a constant factor which we set equal to 0.5, the spherically averaged sensitivity is given by,
\begin{eqnarray}
\delta P_N(k)
&=&  \left[ \sum_{\theta} \left(\frac{1}{P_N(k, \theta)/\sqrt{N_a}} \right)^2
     \right]^{-1/2}
\approx \left[ k^3 \int_{\arccos[\min(\frac{y k}{2 \pi},1)]}^{\arcsin[\min(\frac{k_*}{k}, 1)]} ~ d \theta \,\sin{\theta}
\frac{\epsilon (n(k \sin{\theta}))^2 A_e^3 B^2 t_0^2}{(2 \pi)^2 x^2 y \lambda^6 T_{\rm sys}^4} \right]^{-1/2},
\label{eq:dPk}
\end{eqnarray}
where $k_*$ is the longest transverse wave vector, which corresponds to the maximum baseline length. The lower limit of the integral corresponds to the pixel size.

\section{bispectrum sensitivity}

In this section, we estimate the bispectrum from the thermal noise in a similar way in the previous section. However, we should notice that, because the thermal noise is Gaussian, its bispectrum is actually zero. Nontheless, its statistical fluctuation, that is, its variance is non zero and contributes to the noise to the bispectrum signal. Thus, the calculation in this case is more subtle than that of the power spectrum, although we can use similar techniques as we see below. In \cite{2006MNRAS.366..213S}, an order estimation for the thermal noise bispectrum has been done without considering this fact and also the baseline distribution.

\subsection{covariance of bispectrum}

Remembering the definition of the bispectrum in Eq. (\ref{eq:bs_def}), the covariance of the bispectrum can be defined by,
\begin{eqnarray}
&& {\rm Cov}(B_N(\uvec_1,\uvec_2,\uvec_3) B_N(\uvec_4,\uvec_5,\uvec_6)) D
\nonumber \\
&& = \langle \bigl(\tilde{T}_N(\uvec_1) \tilde{T}_N(\uvec_2) \tilde{T}_N(\uvec_3) - \langle \tilde{T}_N(\uvec_1) \tilde{T}_N(\uvec_2) \tilde{T}_N(\uvec_3) \rangle \bigr)
\bigl(\tilde{T}_N(\uvec_4) \tilde{T}_N(\uvec_5) \tilde{T}_N(\uvec_6) - \langle \tilde{T}_N(\uvec_4) \tilde{T}_N(\uvec_5) \tilde{T}_N(\uvec_6) \rangle \bigr) \rangle
\nonumber\\
&& = \langle \tilde{T}_N(\uvec_1) \tilde{T}_N(\uvec_2) \tilde{T}_N(\uvec_3)
             \tilde{T}_N(\uvec_4) \tilde{T}_N(\uvec_5) \tilde{T}_N(\uvec_6)
     \rangle
\label{eq:CovBB}
\end{eqnarray}
where each bispectrum satisfies the triangular condition ($\uvec_1 + \uvec_2 + \uvec_3 = 0$ and $\uvec_4 + \uvec_5 + \uvec_6 = 0$) and
\begin{eqnarray}
D &=& \delta(\uvec_1-\uvec_4) \delta(\uvec_2-\uvec_5)
      + \delta(\uvec_1-\uvec_4) \delta(\uvec_2-\uvec_6)
      + \delta(\uvec_1-\uvec_5) \delta(\uvec_2-\uvec_4) \nonumber \\
  & & + \delta(\uvec_1-\uvec_5) \delta(\uvec_2-\uvec_6)
      + \delta(\uvec_1-\uvec_6) \delta(\uvec_2-\uvec_4)
      + \delta(\uvec_1-\uvec_6) \delta(\uvec_2-\uvec_5),
\label{eq:D}
\end{eqnarray}
comes from the fact that there is no correlation unless the two triangles, ($\uvec_1,\uvec_2,\uvec_3$) and ($\uvec_4,\uvec_5,\uvec_6$), coincide.

Next, we consider ensemble average of the product of six noise intensities, denoted as $\rm C_B$,
\begin{eqnarray}
C_B(\uvecsix)
&=& \langle \tilde{I}(\uvec_1) \tilde{I}(\uvec_2) \tilde{I}(\uvec_3)
            \tilde{I}(\uvec_4) \tilde{I}(\uvec_5) \tilde{I}(\uvec_6)
    \rangle\nonumber\\
&=& \int d\uvec_1' \int d\uvec_2' \int d\uvec_3'
    \int d\uvec_4' \int d\uvec_5' \int d\uvec_6' 
    \langle \tilde{T}(\uvec_1') \tilde{T}(\uvec_2') \tilde{T}(\uvec_3')
            \tilde{T}(\uvec_4') \tilde{T}(\uvec_5') \tilde{T}(\uvec_6')
    \rangle \nonumber\\
& & \times
    \tilde W(\uvec_1-\uvec_1') \tilde W(\uvec_2-\uvec_2')
    \tilde W(\uvec_3-\uvec_3') \tilde W(\uvec_4-\uvec_4')
    \tilde W(\uvec_5-\uvec_5') \tilde W(\uvec_6-\uvec_6').
\label{eq:C_b}
\end{eqnarray}
To proceed further, we substitute Eq. (\ref{eq:CovBB}) and consider the first term in Eq. (\ref{eq:D}).
\begin{eqnarray}
C_B(\uvecsix)
&=& \int d\uvec_1' \int d\uvec_2' \int d\uvec_3'
    \int d\uvec_4' \int d\uvec_5' \int d\uvec_6' \nonumber\\
& & \times
    {\rm Cov}(B_N(\uvec_1',\uvec_2',\uvec_3') B_N(\uvec_4',\uvec_5',\uvec_6'))
    \delta(\uvec'_1-\uvec'_4) \delta(\uvec'_2-\uvec'_5) \nonumber\\
& & \times
    W(\uvec_1-\uvec_1') W(\uvec_2-\uvec_2') W(\uvec_3-\uvec_3')
    W(\uvec_4-\uvec_4') W(\uvec_5-\uvec_5') W(\uvec_6-\uvec_6')
\nonumber \\
&=& \int d\uvec_1' \int d\uvec_2' \int d\uvec_3' \int d\uvec_6'
    {\rm Cov}(B_N(\uvec_1',\uvec_2',\uvec_3') B_N(\uvec_1',\uvec_2',\uvec_6'))
\nonumber\\
& & \times
    W(\uvec_1-\uvec_1') W(\uvec_2-\uvec_2') W(\uvec_3-\uvec_3')
    W(\uvec_4-\uvec_1') W(\uvec_5-\uvec_2') W(\uvec_6-\uvec_6')
\end{eqnarray}
This is non-zero only when $\uvec_1 \approx \uvec_4$ and $\uvec_2 \approx \uvec_5$ (and then $\uvec_3 \approx \uvec_6$ from the triangular conditions). If these conditions are satisfied,
\begin{eqnarray}
C_B(\uvecsix)
&\approx& {\rm Cov}(B_N(\uvec_1,\uvec_2,\uvec_3) B_N(\uvec_1,\uvec_2,\uvec_3))
          \int d\uvec_1' \int d\uvec_2' \int d\uvec_3' \int d\uvec_6'
\nonumber\\
& & \times
    (W(\uvec_1-\uvec_1'))^2 (W(\uvec_2-\uvec_2'))^2 W(\uvec_3-\uvec_3')
    W(\uvec_6-\uvec_6') \nonumber \\
&=& \bigg(\frac{\lambda^2 B}{A_e}\bigg)^2
    {\rm Cov}(B_N(\uvec_1,\uvec_2,\uvec_3) B_N(\uvec_1,\uvec_2,\uvec_3))
\label{eq:CovB}
\end{eqnarray}
where we used Eq. (\ref{window}) and 
\begin{equation}
\int d\uvec' W(\uvec-\uvec') = 1,
\end{equation}
and assumed ${\rm Cov}(B_N(\uvec_1,\uvec_2,\uvec_3) B_N(\uvec_4,\uvec_5,\uvec_6))$ is approximately constant within the window function. Thus, taking other terms in Eq. (\ref{eq:D}) into account, we have,
\begin{equation}
C_B(\uvecsix)
= D \bigg(\frac{\lambda^2 B}{A_e}\bigg)^2
  {\rm Cov}(B_N(\uvec_1,\uvec_2,\uvec_3) B_N(\uvec_1,\uvec_2,\uvec_3)).
\label{eq:C_B1}
\end{equation}

On the other hand, the product of six noise intensities can also be calculated as follows.
\begin{eqnarray}
C_B(\uvecsix)
&=& \langle \tilde{I}(\uvec_1) \tilde{I}(\uvec_2) \tilde{I}(\uvec_3)
            \tilde{I}(\uvec_4) \tilde{I}(\uvec_5) \tilde{I}(\uvec_6)
    \rangle \nonumber \\
&=& \langle \tilde{I}(\uvec_1) \tilde{I}(\uvec_4) \rangle
    \langle \tilde{I}(\uvec_2) \tilde{I}(\uvec_5) \rangle
    \langle \tilde{I}(\uvec_3) \tilde{I}(\uvec_6) \rangle
    + ({\rm 5~permutations }) \nonumber\\
&=& D \langle \tilde{I}(\uvec_1) \tilde{I}(\uvec_1) \rangle
      \langle \tilde{I}(\uvec_2) \tilde{I}(\uvec_2) \rangle
      \langle \tilde{I}(\uvec_3) \tilde{I}(\uvec_3) \rangle \nonumber\\
&=& D \Bigl(\frac{\lambda^{2} B T_{\rm sys}}{A_{e}}\Bigr)^6
    \frac{1}{B^3 t_{\uvec_1} t_{\uvec_2} t_{\uvec_3}},
\label{eq:C_B2}
\end{eqnarray}
where we used Wick theorem \citep{2009A&A...508.1193J} in the second equality and Eq. (\ref{eq:noise_pix}) in the last equality.

Thus, from Eqs. (\ref{eq:C_B1}) and (\ref{eq:C_B2}), we obtain,
\begin{equation}
{\rm Cov}(B_N(\uvec_1,\uvec_2,\uvec_3) B_N(\uvec_1,\uvec_2,\uvec_3))
= \bigg(\frac{A_e}{\lambda^2 B}\bigg)^2
  \Bigl(\frac{\lambda^{2} B T_{\rm sys}}{A_e}\Bigr)^6
  \frac{1}{B^3 t_{\uvec_1}t_{\uvec_2}t_{\uvec_3}}.
\end{equation}
Converting the argument from $\uvec$ to $\kvec$, we finally obtain,
\begin{eqnarray}
{\rm Cov}(B_N(\kvec_1,\kvec_2,\kvec_3) B_N(\kvec_1,\kvec_2,\kvec_3))
&=& \bigg(\frac{x^2 y}{B}\bigg)^4
    {\rm Cov}(B_N(\uvec_1,\uvec_2,\uvec_3) B_N(\uvec_4,\uvec_5,\uvec_6))
\nonumber \\
&=& \bigg(\frac{x^2 y \lambda^2}{A_e}\bigg)^4
    \frac{T^6_{\rm sys}}{B^3 t_{\kvec_1} t_{\kvec_2} t_{\kvec_3}}.
\label{eq:CovBB}
\end{eqnarray}
This equation corresponds to Eq. (\ref{eq:noise_pix}) for the power spectrum, if we substitute Eq. (\ref{eq:tu}).

\subsection{spherical average}

In this subsection, we take a sum of the noise bispectrum over spherical shell as we did for the power spectrum in the previous section. However, the situation is much more complicated in the case of bispectrum, because $|\kvec_1|, |\kvec_2|$ and $|\kvec_3|$ can be all different with each other in general so that we must consider two spherical shells with the radius $|\kvec_1|$ and $|\kvec_2|$, while $|\kvec_3|$ is determined by the triangular condition, $\kvec_1 + \kvec_2 + \kvec_3 = 0$. In this paper, we calculate the noise bispectrum for equilateral type ($|\kvec_1| = |\kvec_2| = |\kvec_3|$) and isosceles type ($|\kvec_2| = |\kvec_3|$) and define $K \equiv |\kvec_1|$ and $k \equiv |\kvec_2| = |\kvec_3|$.

First, as in the case of the power spectrum, $\kvec_1$ can run over a spherical shell with radius $k$ which can be parametrized by two of the spherical coordinate of $\kvec_1$, $(\theta_1,\phi_1)$. Further, for a fixed $\kvec_1$, there is a rotational degree of freedom for $\kvec_2$ with respect to $\kvec_1$, which is denoted by an angle $\alpha$ with $0 \leq \alpha < 2 \pi$. Thus, we need to integrate the covariance matrix in Eq. (\ref{eq:CovBB}) with respect to $\theta_1$, $\phi_1$ and $\alpha$. Noting that the covariance matrix does not depend on $\phi_1$, the weight of the integration, which corresponds to Eq. (\ref{pixnum}), is given by
\begin{eqnarray}
N_a
= \left[ 2 \pi \sin{\theta_1} K^2 \Delta K \Delta \theta_1 \frac{V}{(2\pi)^3}
  \right]
  \times
  \left[ k^2 \sin{\theta_2} \sin{\gamma} ~ \Delta k \Delta \theta_2
         \Delta \alpha \frac{V}{(2\pi)^3}
  \right].
\end{eqnarray}
where the first factor comes from the sum for $\kvec_1$ over the spherical shell and the second factor takes the rotational degree of freedom of $\kvec_2$ for each $\kvec_1$ into account. Here $\theta_2$ is the polar angle of $\kvec_2$ and $\gamma$ is the angle $\partial \kvec_2/\partial \alpha$ and $\partial \kvec_2/\partial \theta_2$. $\Delta \theta_2$ is the width of the annulas of $\kvec_2$ when $\kvec_1$ is fixed, which we set equal to the resolution in Fourier space, $2 \pi/V^{1/3}$.

It is convenient to express $\theta_2$ by $\theta_1$, $\alpha$ and the angle between $\kvec_1$ and $\kvec_2$ denoted as $\beta$. Noting $\kvec_2$ can be express as
\begin{equation}
\kvec_2 = k (\cos{\theta_1} \cos{\alpha} \sin{\beta} + \sin{\theta_1} \cos{\beta}, \sin{\alpha} \sin{\beta}, - \sin{\theta_1} \cos{\alpha} \sin{\beta} + \cos{\theta_1} \cos{\beta}),
\end{equation}
we obtain,
\begin{equation}
\cos{\theta_2}
= - \sin{\theta_1} \cos{\alpha} \sin{\beta} + \cos{\theta_1} \cos{\beta}
\end{equation}
Then, setting $\Delta k = \epsilon k$ and $\Delta K = \epsilon K$, the bispectrum variance due to the thermal noise is written by an integrate with respect to $\theta_1$ and $\alpha$,
\begin{eqnarray}
\delta B_N(k,K,\beta)
&=& \left[ \sum_{\theta}  \sum_{\alpha}
    \left(\frac{1}{\sqrt{N_a}} \sqrt{{\rm Cov}(B_1B_2)(k, K, \theta_1, \alpha)}     \right)^{-2} \right]^{-1/2} \nonumber \\
&=& \frac{(2\pi)^{\frac{5}{2}}}{\sqrt{\Delta \theta_2} k K^{3/2} \epsilon}
    \bigg(\frac{x^2y \lambda^2}{A_e}\bigg)
    \bigg(\frac{T^2_{\rm sys} \lambda^2}{A_e B t_0}\bigg)^{\frac{3}{2}}
\nonumber \\
& & \times \left[\int {d\theta_1} \int {d\alpha} \sin{\theta_1} \sin{\theta_2}
                 \sin{\gamma(\theta_1,\alpha)}
                 ~ n(\kvec_1) n(\kvec_2) n(\kvec_3)
           \right]^{-\frac{1}{2}}.
\end{eqnarray}
This is a general expression for isosceles-type bispectrum. For equilateral type, we just set $K = k$ and $\beta = 2 \pi/3$.

\subsection{estimation of noise bispectrum}

To calculate the bispectrum sensitivity, we need the number density of baselines on uv-plane. In this paper, we consider expanded MWA, LOFAR and SKA. The expanded MWA will have 500 antennae within a radius of 750 m with $r^{-2}$ distribution \citep{2006ApJ...638...20B}. LOFAR has 24 antennae within a radius of 2000 m with $r^{-2}$ distribution \citep{2013A&A...556A...2V}. HERA has 547 antennae within 200 m with constant distribution \citep{2014ApJ...782...66P}. SKA will have 466 antennae within 600 m with $r^{-2}$ distribution, 670 antennae within 1000 m, 866 antennae within 3000 m \citep{SKA}. For simplicity, we assume that the antennae density is constant between 600 m to 1000 m and 1000 m to 3000 m, respectively. We list parameters in table \ref{table:parameter}. Further, we assume $t_0 = 1000~{\rm hour}$ for the total observing time and $6~{\rm MHz}$ bandwidth.

\begin{table}[tb]
\begin{center}
\begin{tabular}{|c|c|c|c|c|c|}
\hline
redshift             & 8   & 10  & 12   & 17   & $N_{\rm station}$ \\ \hline
frequency [MHz]      & 158 & 129 & 109  & 79   &     \\ \hline
$T_{\rm sys}~[K]$    & 440 & 600 & 1000 & 1900 &     \\ \hline
$A_e~[m^2]$ (MWA)    & 14  & 18  & 18   & 18   & 500 \\ \hline
$A_e~[m^2]$ (LOFAR)  & 512 & 600 & 900  & 900  & 24  \\ \hline
$A_e~[m^2]$ (HERA)   & 68  & 106 & 154  & 154  & 547 \\ \hline
$A_e~[m^2]$ (SKA)    & 462 & 728 & 962  & 962  & 866 \\ \hline
\end{tabular}
\caption{Parameters for telescopes: $T_{\rm sys}$ is system temperature, $A_e$ is effective area of a station and $N_{\rm station}$ is the number of stations.}
\label{table:parameter}
\end{center}
\end{table}

For comparison, we show the bispectrum of 21cm signal from the epoch of reionization, using a public code, 21cmFAST \citep{2011MNRAS.411..955M}. This is based on a semi-analytic model of reionization and we can obtain 3D brightness temperature maps at arbitrary redshifts. We set the simulation box to $(200~{\rm Mpc})^3$ with $300^3$ grids and take a fiducial set of model parameters as $(\zeta, \zeta_X, T_{\rm vir}, R_{\rm mfp}) = (31.5, 10^{56}/M_{\odot}, 10^4~{\rm K}, 30~{\rm Mpc}$). Here, $\zeta$ is the ionizing efficiency, $\zeta_X$ is the number of X-ray photons per solar mass, $T_{\rm vir}$ is the minimum virial temperature of halos which host stars and $R_{\rm mfp}$ is the mean free path of ionizing photons. {We also estimate the sample variance of the bispectrum by calculating the average and variance from 19 brightness-temperature maps with different realizations of the initial condition.}

\begin{figure}[t]
\centering
\includegraphics[width=11.5cm,bb=0 0 330 272]{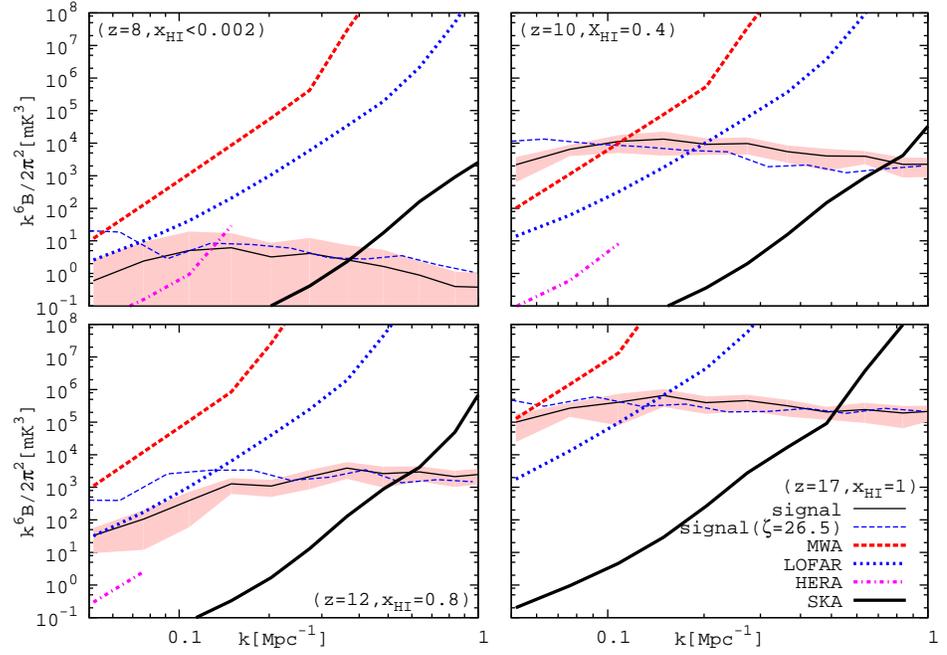}
\caption{Comparison of equilateral-type bispectrum of 21cm signal {of the fiducial model with sample variance (thin dashed line and shaded area), signal of a variant model with $\zeta = 26.5$ (thin dot dashed line)} and thermal noise  of MWA (thick dashed), LOFAR (thick dotted), HERA(thick dot dashed) and SKA (thick solid) at $z = 8,10,12$ and $17$, {Here $x_{\rm HI}$ is the neutral fraction of the fiducial model at each redshift}.}
\label{eqi1}
\end{figure}

In Fig. \ref{eqi1}, we compare the equilateral-type bispectrum signal with thermal noise at $z = 8,10,12$ and $17$. {Here the sample variance for the fiducial model and the average signal for a variant model with $\zeta = 26.5$ are also shown for reference. The ionization is less effective for this variant model so that the reionization proceeds slowly compared with the fiducial model.} Generally, the noise increases toward smaller scales which reflects the deficiency of longer baselines. On the other hand, the sensitivity for larger scales are limited by the survey volume. We see the signals are larger than SKA noise for $k \lesssim 0.3~{\rm Mpc}^{-1}$ at all redshifts. However, the thermal noise dominates over the signal for the expanded MWA at almost all scales and redshifts, while the bispectrum may be observable for large scales $k \lesssim 0.05~{\rm Mpc}^{-1}$ at $z = 10$. LOFAR has better sensitivity and the signal will be observable at scales with $k \lesssim 0.1~{\rm Mpc}^{-1}$ at $z = 10$ and $17$. Here it should be noted that the bispectrum signal has several peaks as a function of redshift and they are at $z = 10$ and $17$ \citep{shi-P}. The peak redshifts depend on the specific values of the model parameters and observation of them will give us information on the process of reionization. Thus, it is expected that LOFAR is enough sensitive to detect the bispectrum at the peak redshifts for large scales. {On the other hand, due to the short baselines of HERA, equilateral triangle does not exist in the $(u,v,\eta)$ space of HERA for large $k$, while it has a sufficient sensitivity for small $k$.}

\begin{figure}[t]
\centering
\includegraphics[width=11.5cm,bb=0 0 330 272]{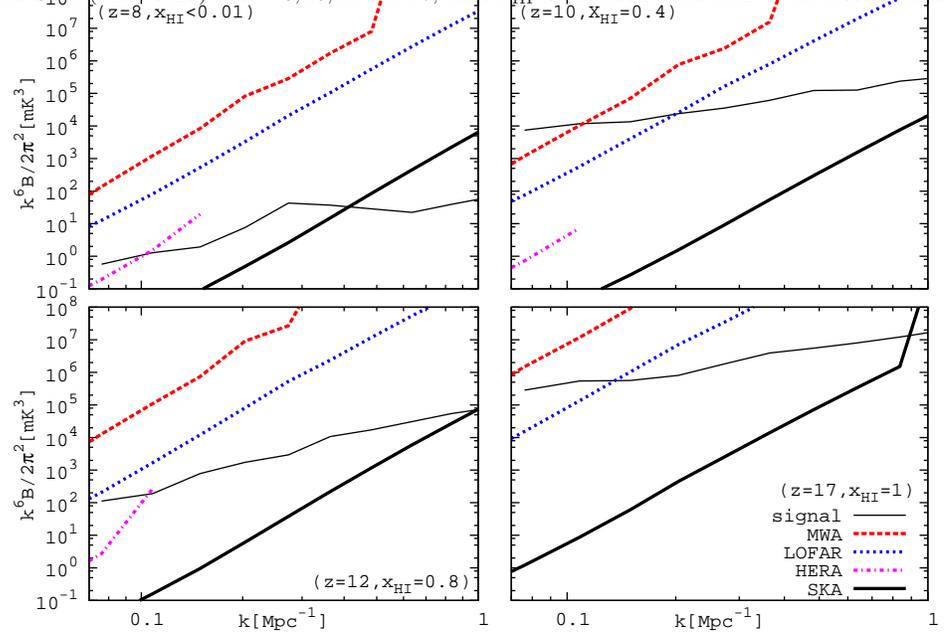}
\caption{Comparison of isosceles-type bispectrum of 21cm signal (thin dashed line) and thermal noise  of MWA (thick dashed), LOFAR (thick dotted), HERA(dot dashed) and SKA (thick solid) at $z = 8,10,12$ and $17$. Here $x_{\rm HI}$ is the neutral fraction of standard model at each redshift. Here one of the wavenumber $K$ is fixed to $0.06~{\rm Mpc}^{-1}$.}
\label{iso1}
\end{figure}

The isosceles-type bispectrum with $K = 0.06~{\rm Mpc}^{-1}$ bispectra are plotted in Fig. \ref{iso1}. The behavior and relative amplitudes of the signal and noise are very similar to the case of the equilateral type but SKA is more sensitive at smaller scales.

{Finally, we calculate the total signal-to-noise ratio for equilateral and isosceles types, considering a $k$ range from $5.0 \times 10^{-2}~{\rm Mpc^{-1}}$ to $1.0~{\rm Mpc^{-1}}$ in Figs. \ref{eqi1} and \ref{iso1}, respectively. Here the total signal-to-noise ratio is calculated by summing the square of signal-to-noise ratio over $k$ bins and taking its square root. The result is shown in the Table \ref{table:signal_to_noise}.}

\begin{table}[tb]
\begin{center}
\begin{tabular}{|c|c|c|c|c|c|}
\hline
redshift &~~~8~~~~&~~~10~~~~&~~~12~~~~&~~~17~~~~\\ \hline
MWA(equ) & $5.4\times10^{-2}$ & $2.2\times10^{1}$ & $3.1\times10^{-2}$ & $8.0\times10^{-1}$  \\ \hline
LOFAR(equ) & $3.5\times10^{-1}$ & $1.9\times10^{2}$ & $1.3$ & $6.0\times10^{1}$  \\ \hline
HERA(equ) & $2.6\times10^{1}$ & $2.4\times10^{4}$ & $1.2\times10^{2}$ & $3.7\times10^{3}$  \\ \hline
SKA(equ) & $1.7\times10^{3}$ & $1.9\times10^{6}$ & $1.1\times10^{4}$ & $5.7\times10^{5}$  \\ \hline
MWA(iso) & $4.3\times10^{-3}$ & $6.4$& $8.9\times10^{-3}$ & $2.0\times10^{-1}$  \\ \hline
LOFAR(iso) & $4.8\times10^{-2}$ & $1.0\times10^{2}$ & $5.5\times10^{-1}$ & $1.8\times10^{1}$  \\ \hline
HERA(iso) & $3.0$ & $2.4\times10^{4}$ & $4.0\times10^{1}$ & $0.0$  \\ \hline
SKA(iso) & $1.1\times10^{2}$ & $4.3\times10^{5}$ & $2.3\times10^{3}$ & $1.0\times10^{5}$  \\ \hline
\end{tabular}
\caption{{Total signal-to-noise ratios of bispectra of equilateral (equ) and isosceles (iso) types for each array and redshift.}}
\label{table:signal_to_noise}
\end{center}
\end{table}

\section{summary and discussion}

In this paper, we esimated the bispectrum of thermal noise for redshifted 21cm signal observation for Epock of Reionization by extending the formalism of the noise power spectrum estimation given by \cite{2006ApJ...653..815M}. Because thermal noise was assumed to be Gaussian, the ensamble average of the bispectrum vanishes and its variance contributes to the noise to the bispectrum signal. We developed a formalism to calculate the noise bispectrum for an arbitrary triangle, taking the array configuration into account. We applied it to the cases with equilateral and isosceles triangles and estimated the noise bispectrum for expanded MWA, LOFAR and the SKA. Consequently, it was found that the SKA has enough sensitivity for $k \lesssim 0.3~{\rm Mpc}^{-1}$ for both types of triangles. On the other hand, LOFAR will have sensitivity for the peaks of the bispectrum as a function of redshift. The expanded MWA has even less sensitivity but it will be possible to put a meaningful constraints on model parameters which induce larger signals than those with the parameters used in this paper.

Not only the themal noise but signal of bispectrum depend on the configuration of the triangle of three wave numbers. It is possible that the signal bispectrum has a large amplitude for a specific configuration of the triangle and observation may become easier in that case. An investigation of the details of the bispectrum signal and comparison with noise bispectrum will be presented elsewhere \citep{shi-B}.

Actually, thermal noise is just one of many obstacles for the observation of 21cm signal. Other serious sources of noise are Galactic and extragalactic foreground and sample variance, and the foreground emission has not been well understood even for power spectrum. {It may be helpful to consider the "EoR window" for bispectrum as in the case of power spectrum \citep{2014ApJ...782...66P}. It would turn out that small-k modes should be discarded to avoid foreground and, if this is the case, the total signal-to-noise ratios in Table \ref{table:signal_to_noise} are overestimation. Further, for a practical application of bispectrum analysis to survey data, we need to take survey geometry into account \citep{2000ApJ...544..597S,2001ApJ...546..652S,2014arXiv1407.5668G}.}  Nevertheless, the observation of the bispectrum is very important because 21cm signal from Epoch of Reionization is highly non-Gaussian so that the bispectrum will give us enormous information complementary to the power spectrum.

\section*{Acknowledgement}
This work is supported by Grant-in-Aid from the Ministry of Education, Culture, Sports, Science and Technology (MEXT) of Japan, Nos. 24340048 and 26610048(K.T), No. 25-3015(H.S.), No. 26800104(H.N.) and No. 25610043(H.I.).

\end{document}